\documentclass{aastex61}
\usepackage{graphicx,epstopdf,epsfig}
\usepackage{natbib}

\shortauthors{Mazumder et al.}

\begin{document}

\title{The Association of Filaments, Polarity Inversion Lines, and Coronal Hole Properties with the Sunspot Cycle: An Analysis of the McIntosh Database}

\author{Rakesh Mazumder}
\affiliation{Center of Excellence in Space Sciences India, Indian Institute of Science Education and Research Kolkata, Mohanpur 741246, West Bengal, India}

\author{Prantika Bhowmik}
\affiliation{Center of Excellence in Space Sciences India, Indian Institute of Science Education and Research Kolkata, Mohanpur 741246, West Bengal, India}

\author{Dibyendu Nandy}
\correspondingauthor{Dibyendu Nandy}
\email{dnandi@iiserkol.ac.in}
\affiliation{Center of Excellence in Space Sciences India, Indian Institute of Science Education and Research Kolkata, Mohanpur 741246, West Bengal, India}
\affiliation{Department of Physical Sciences, Indian Institute of Science Education and Research Kolkata, Mohanpur 741246, West Bengal, India}

\begin{abstract}

Filaments and coronal holes, two principal features observed in the solar corona are sources of space weather variations. Filament formation is closely associated with polarity inversion lines (PIL) on the solar photosphere which separate positive and negative polarities of the surface magnetic field. The origin of coronal holes is governed by large-scale unipolar magnetic patches on the photosphere from where open magnetic field lines extend to the heliosphere. We study properties of filaments, PILs and coronal holes in solar cycles 20, 21, 22 and 23 utilizing the McIntosh archive. We detect a prominent cyclic behavior of filament length, PIL length, and coronal hole area with significant correspondence with the solar magnetic cycle. The spatio-temporal evolution of the geometric centers of filaments shows a butterfly-like structure and distinguishable pole-ward migration of long filaments during cycle maxima. We identify this rush to the poles of filaments to be co-temporal with the initiation of polar field reversal as gleaned from Mount Wilson and Wilcox Solar Observatory polar field observations and quantitatively establish their temporal correspondence. We analyze the filament tilt angle distribution to constrain their possible origins. Majority of the filaments exhibit negative and positive tilt angles in the northern and the southern hemispheres, respectively -- strongly suggesting that their formation is governed by the overall large-scale magnetic field distribution on the solar photosphere and not by the small-scale intra-active region magnetic field configurations. We also investigate the hemispheric asymmetry in filaments, PILs, and coronal holes. We find that the hemispheric asymmetry in filaments and PILs are positively correlated -- whereas coronal hole asymmetry is uncorrelated -- with sunspot area asymmetry. 

\end{abstract}


\section{Introduction}
Solar filaments are one of the most common large-scale features observed in the solar corona. Filaments are made of dense chromospheric plasma which is much cooler as compared to 1 MK hot corona; thus they appear as dark elongated structures against the bright solar disk when observed in H-$\alpha$ absorption line. The same structures look brighter on the solar limb in emission line and are known as prominences. Hydrodynamic instabilities can trigger eruptions of filaments, producing solar flares and coronal mass ejections which are hazardous for space weather \citep{2000ApJ...537..503G,2003ApJ...586..562G,2004ApJ...614.1054J}. Therefore observational as well as theoretical studies of filaments have attracted much attention in the context of space weather research. Though filaments are coronal features, their properties depend on the magnetic environment of the solar surface (i.e., the photosphere). They always appear in the vicinity of the polarity inversion lines (PILs) which separate opposite polarities of the photospheric magnetic field \citep{1998SoPh..182..107M,1972RvGSP..10..837M,1983SoPh...85..215M}. Thus filaments inherit diverse signatures of the temporal and spatial variations of the small- and large-scale surface magnetic fields -- establishing them as a crucial component for studying solar variability. Moreover, analysis of filament data spanning over several solar cycles can reveal various aspects of the long-term modulation of the solar magnetic field. 

Systematic observations of filaments (in H$_{\alpha}$ line) started long back, in 1915, at Kodaikanal Observatory (India); and 1919, at the Kislovodsk Mountain Astronomical Station of the Main (Pulkovo) Astronomical Observatory of Russian Academy of Sciences. Utilizing these multiple databases, researchers \citep{2015ApJS..221...33H,2003NewA....8..655L,2008JGRA..11311108L,2017ApJ...849...44C} have identified a cyclic time variation in the total filament number (also, in their total length, \citealt{2016SoPh..291.1115T}) -- quite similar to the solar cycle. In general, filaments are distributed over all possible latitudes (from the equator to the poles) and longitudes on the solar surface throughout a solar cycle. However, \cite{2015ApJS..221...33H}, \cite{2016SoPh..291.1115T}, and \cite{2017ApJ...849...44C} reported that the temporal variation of the latitudinal positions of filaments' geometric centers forms a butterfly-like pattern with a spread much larger as compared to the sunspot butterfly diagram (which is usually confined within $\pm$45$^\circ$ latitudes). Their work also includes statistical analysis of filaments' tilt angle distribution. Like many other solar activities, filament formation is not uniform in the two solar hemispheres. \cite{2003NewA....8..655L,2010NewA...15..346L,2015RAA....15...77K,2015ApJS..221...33H} investigated north-south asymmetry of filaments during different phases of the solar cycle. 

Apart from solar filaments, another large-scale coronal feature is coronal holes through which unipolar open magnetic field lines with roots in the solar photosphere extend to the heliosphere. Chromospheric ionized plasma particles gyrate along these lines and propagate in the upper atmosphere quite easily, making coronal holes a potential source of fast solar wind. Geomagnetic storms are primarily caused by this fast solar wind \citep{2000ApJ...537..503G,2003ApJ...586..562G,2012A&A...542A..52Z}, and hence, prediction of the solar wind properties is a key to space weather research applications. Low temperature and density of coronal holes in comparison to the surroundings make them appear as dark patches in X-ray \citep{1975SoPh...42..135T} and EUV \citep{1972ApJ...176..511M} images. Polar coronal holes have been studied using He II 30.4 nm \citep{1977SoPh...51..377B}, He I 1083 nm \citep{1980SoPh...65..229S,1984SoPh...92..109W,1998SoPh..177..375F,2002SoPh..211...31H} and Fe XIV 530.3 nm \citep{1981SoPh...70..251W,2001SoPh..203...27M} coronal emission lines, K-coronograph observation \citep{1994SoPh..154..377B}, X-ray and EUV observation \citep{1984SoPh...92..109W,1978SoPh...56..161B} and potential field source surface (PFSS) extrapolation modeling utilizing photospheric magnetic field data \citep{1996Sci...271..464W,1998SoPh..179..223B,1999SoPh..187..185O}. Recently, \cite{2014ApJ...783..142L, 2017SoPh..292...18L} have used EIT/SOHO and SDO/AIA data to study the signed and the unsigned magnetic flux confined in the coronal holes during solar cycles 23 and 24. They have compared these observational results with theoretically calculated magnetic flux associated with those coronal holes by utilizing PFSS modeling on proper surface magnetic maps. They have also reported the presence of asymmetry in area evolution of coronal holes in the two hemispheres. In a similar context, \cite{1999SoPh..187..185O,2009ApJ...707.1372W,2015NatCo...6E6491M,2016SSRv..tmp...21W} have studied the coronal hole area variation during solar cycles and found a butterfly-like structure in the time--latitude plot of coronal holes' centroids. In a recent work, \cite{2018AJ....155..153K} studied the lifetimes and propagation characteristics of low latitude coronal holes in longitude over several solar rotations using the SDO and Stereo data for a period starting from February, 2011 to July, 2014.

In this paper, we analyze observational data preserved in the McIntosh archive (McA:\citealt{2016AGUFMSH11A2220G,SWE:SWE20537}) containing positional information of various observational features such as filaments, PILs, coronal holes, sunspots and plages during solar cycles 20 to 23. We mainly focus on the variation of filament length and its hemispheric asymmetry. We further analyze the latitudinal and tilt angle distributions of filaments during this period. We additionally explore various properties of PILs and coronal holes and establish their relationship to the solar magnetic activity. The paper is organized in the following way: in section 2, we describe the source of the data; results of our analyses are presented in section 3; the last section is dedicated to conclusions from our analyses. 

\section{Data Description}
Patrick McIntosh, a scientist at NOAA's Space Environment Center in Boulder, started creating hand-drawn Carrington maps using both satellite and ground-based observations from April 1967 to July 2009. A Carrington map contains positional information of different features observed on the solar photosphere (e.g., sunspots, PILs, and plages) and corona (e.g., filaments and coronal holes). By assembling these maps, he developed a unique database capturing the long-term solar activity during solar cycles 20--23. However, there are three big data-gaps: the first one during June 1974 -- July 1978, the second one during October 1991 -- January 1994, and the last one during April 1994 -- May 1996. In McIntosh archive project (a Boston College/NOAA/NCAR collaboration, funded by the NSF), all hand-drawn maps have been scanned, digitized and archived at NOAA/NCEI and made available both as images and `fits' format\footnote{\url{https://www2.hao.ucar.edu/mcIntosh-archive/four-cycles-solar-synoptic-maps}}. For our analysis, we have used level 3 fits files from the McIntosh archive and extracted the data associated with the relevant features present in those Carrington maps.

\section{results and discussion}

\subsection{Properties of filaments}
Filament formation is strongly associated with the magnetic field distribution on the solar surface. Thus variation in the appearance of filaments can delineate the magnetic activity of the Sun. In this work, we analyze 442 Carrington map images from April 1967 to July 2009 distributed over approximately 36 years, covering mostly the declining phases of solar cycle 20, cycle 21, some part of cycle 22, full solar cycle 23, and the beginning of cycle 24. A total of 67373 filaments are detected from these 442 Carrington maps with their indices varying from 1520 to 2086. In order to calculate the length of an individual filament present in any Carrington map, we first identify the boundary pixels of the filamentary structure and then measure the associated perimeter by using the following relation,
\begin{equation}
 L = \sum_{n} \sqrt{R_{\odot}^{2}\delta\theta^{2}+R_{\odot}^{2}Cos^2 \theta \delta\phi^{2}}
\end{equation}
\noindent where L is the length of the filament's perimeter, $R_{\odot}$ is radius of the Sun, the symbols $\theta$ and $\phi$ represent the latitude and the longitude of a particular pixel and $n$ being the total number of pixels associated with the filament's perimeter. The quantities $\delta \theta$ and $\delta\phi$ are latitudinal and longitudinal differences between two adjacent pixels, respectively. We consider the length of any individual filament to be half of its perimeter. While fitting a Gaussian to the filaments' length distribution, we find the mean and the standard deviation ($\sigma$) to be 40 Mm and 24 Mm, respectively, with a very long tail extending up to 1894 Mm. We investigate different characteristics of these filaments and explore the possible connection to the solar magnetic cycle.

\subsubsection{Latitudinal distribution of filaments}
\begin{figure}
\includegraphics[scale=0.9,angle=90]{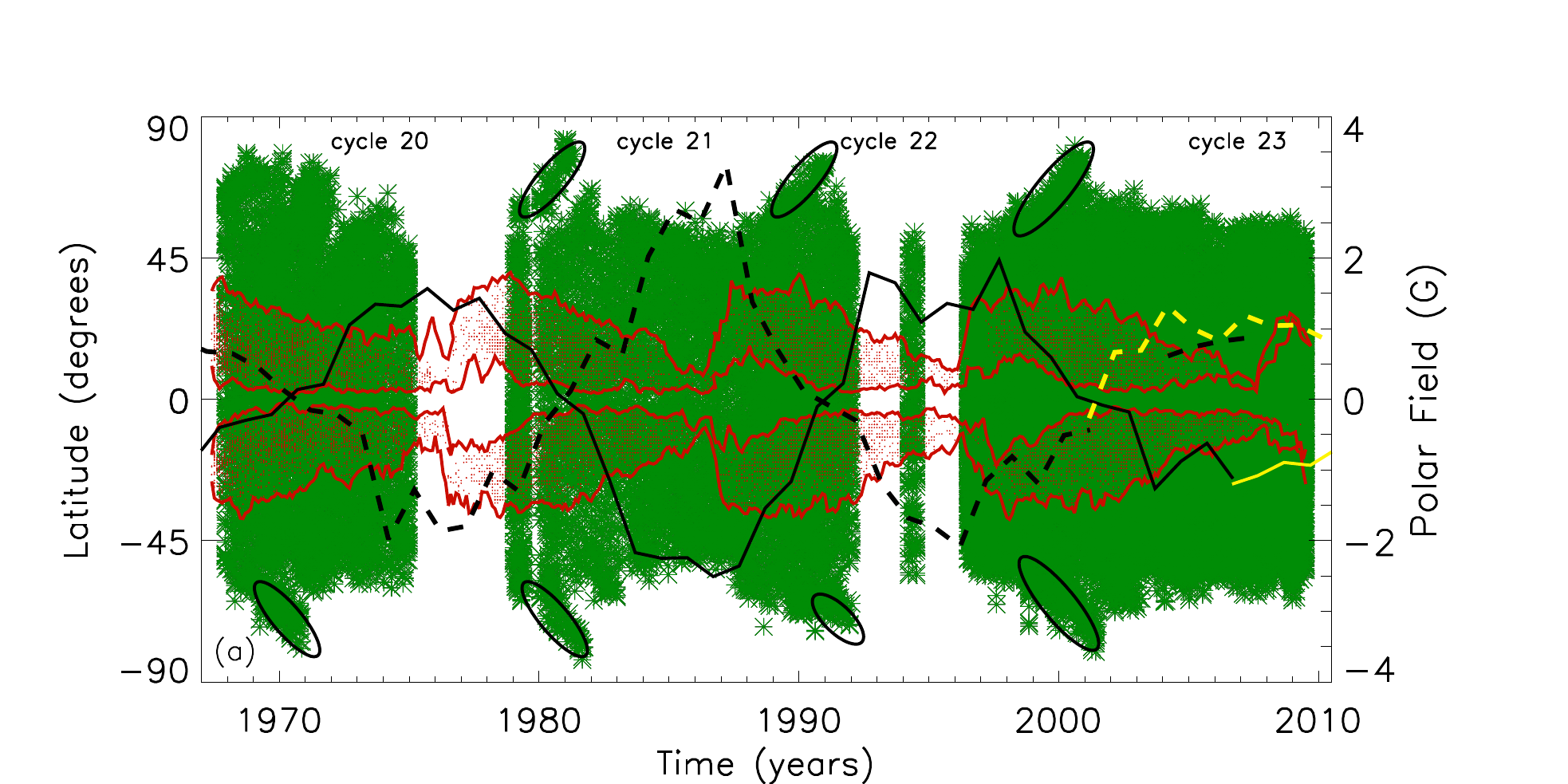}
\includegraphics[scale=0.9,angle=90]{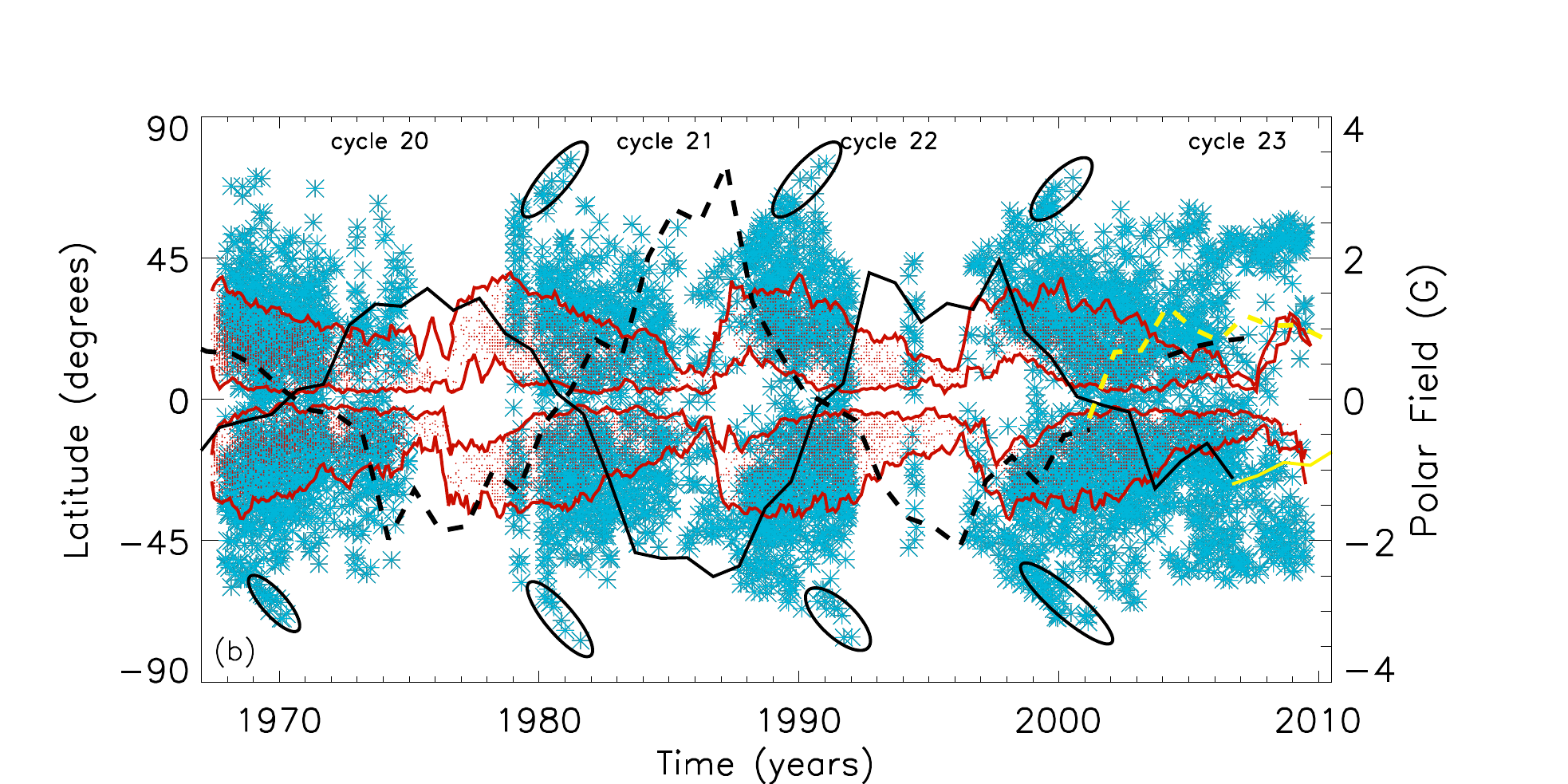}
\caption{The temporal evolution of filament's latitudinal distribution for majority of filaments is depicted in panels (a) and that for 'very long' filaments is depicted in panel (b). In panel (a), the green asterisks represent geometric centers of individual average filaments (length $<$ 88 Mm). The red contours represent the latitudinal coverage by sunspot activity belts. The geometric centers of sunspots are marked by red dots. The solid and dashed black lines show the temporal variation of polar magnetic field (in the northern and the southern hemispheres, respectively), with the data taken from the Mount Wilson Observatory (MWO). The solid and dashed yellow lines depict same while using the data taken from the Wilcox Solar Observatory (WSO). The salient feature of filaments, `rush to the pole', are identified and highlighted by encircling black ellipses. It is evident that the timing of polar field reversal and polar rush of long filaments coincide with each other. In panel (b) the blue asterisks correspond to the latitudinal positions of the geometric centers of very long filaments (length $>$ 184 Mm). The Polar field variation and sunspots' activity belts are depicted following the same convention as used in panel (a).}
\label{fig1}
\end{figure}

Filaments are not uniformly distributed over the solar surface. While there is no prominent longitudinal preference for filament formation, we find a discernible pattern in their latitudinal position during a solar cycle. Both the Figures \mbox{\ref{fig1}}(a) and (b) represent the spatio-temporal variation of filaments' geometric centers (corresponding to different length scales), showing a wider spread in latitude (-$80^{\circ}$ to +$80^{\circ}$ degrees) in comparison to the sunspot butterfly diagram (denoted by red dots within red contour lines). The sunspot data is taken from the Royal Greenwich Observatory (RGO) and US Air Force (USAF) - Solar Optical Observing Network (SOON) database. Figure \mbox{\ref{fig1}}(a) depicts the spatio-temporal distribution of filaments with length varying from 2 Mm (smallest) to 88 Mm ($\approx$ mean+2$\sigma$) capturing the distribution of the majority (i.e., 76$\%$) of all filaments. While we consider the geometric centers of the very long filaments with length $>$ 184 Mm (i.e., mean+6$\sigma$), a more apparent butterfly-like structure emerges [see, Figure \mbox{\ref{fig1}}(b)], although they form a small subset of all filaments (7.7$\%$). 

At higher latitudes, filaments display a pole-ward migration [see, the structures encircled by black ellipses, which are more prominent in Figure \mbox{\ref{fig1}(b)}] during the solar maxima of cycle 20 (the year 1971), cycle 21 (the year 1981), cycle 22 (the year 1990) and cycle 23 (the year 2000). This observational characteristic of filaments is known as `rush to the poles' \citep{1982SoPh...79..231T,2016SoPh..291.1115T,2017ApJ...849...44C}. The presence of similar features was reported in an analysis of the poleward-most filaments using the McIntosh Archive (only, cycle 23) performed by \cite{2016AGUFMSH11A2220G}. In the same figures, we have represented the variation of polar field observed by Mount Wilson Observatory (MWO) (north: solid black line and south: dashed black line) and Wilcox Solar Observatory (WSO) (north: solid yellow line and south: dashed yellow line). It is clearly visible that the reversal of polar field begins when the large-scale magnetic structures hosting filaments (see, the `tongue'-like features) reach near the polar region in both the hemispheres. This result reconfirms earlier reports that the beginning of these `rush to the poles' can be treated as an indicator of the imminent reversal of the polar field \citep{1989SoPh..123..367M,1997SoPh..170..411A,2006PASJ...58...85S,2008JGRA..11311108L,2015ApJS..221...33H,2006PASJ...58...85S,2016ApJ...823L..15G}.

\begin{figure}[hb!]
\centering
\includegraphics[scale=0.45,angle=0]{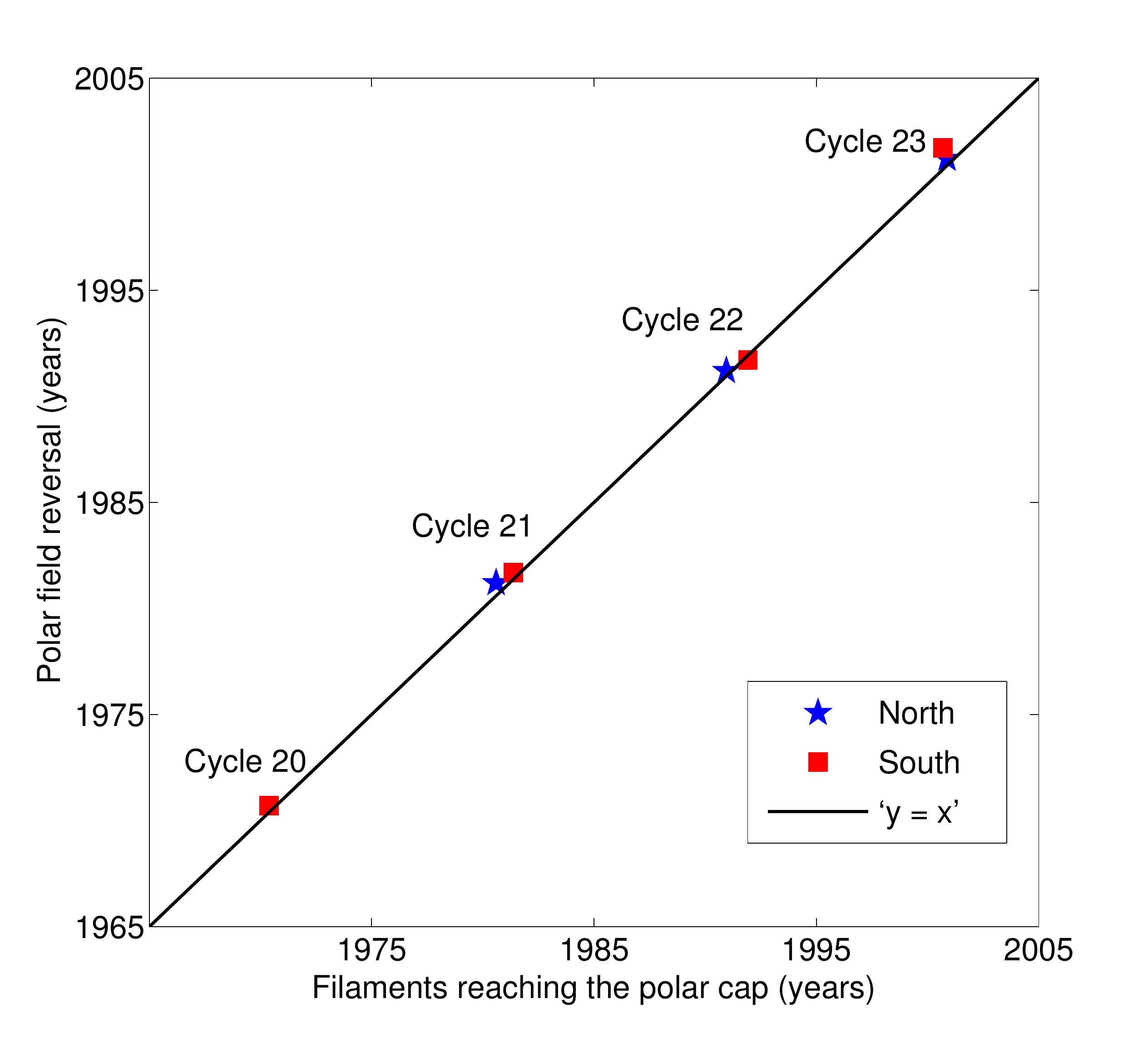}
\caption{Timing of polar field reversal and the time when filaments reach the polar cap region are compared in the above figure. The red squares and blue stars represent the timing in the southern and the northern hemispheres, respectively. The straight line `$y=x$' (depicted by a black solid line) passes through the data points and validates our conjecture that arrival of `rush to the pole' features associated with filaments near the polar cap regions indicates the upcoming polar field reversal.}
\label{fig2}
\end{figure}

We further compare the timing of filaments reaching the polar caps (regions beyond $\pm 75^{\circ}$) with the reversal of polar field in both the hemispheres and investigate the degree of their temporal conjunction. In Figure \ref{fig2}, the data points denoted by blue stars and red squares (for the northern and the southern hemispheres, respectively) represent the relevant timings quantitatively. The overlap of `$y=x$' line over the data points confirms our conjecture that `rush to the poles' is a marker of the imminent reversal of the polar field.

\begin{figure}[!tbp]
  \centering
  \begin{minipage}[b]{0.4\textwidth}
    \includegraphics[angle=90,width=1.2\textwidth]{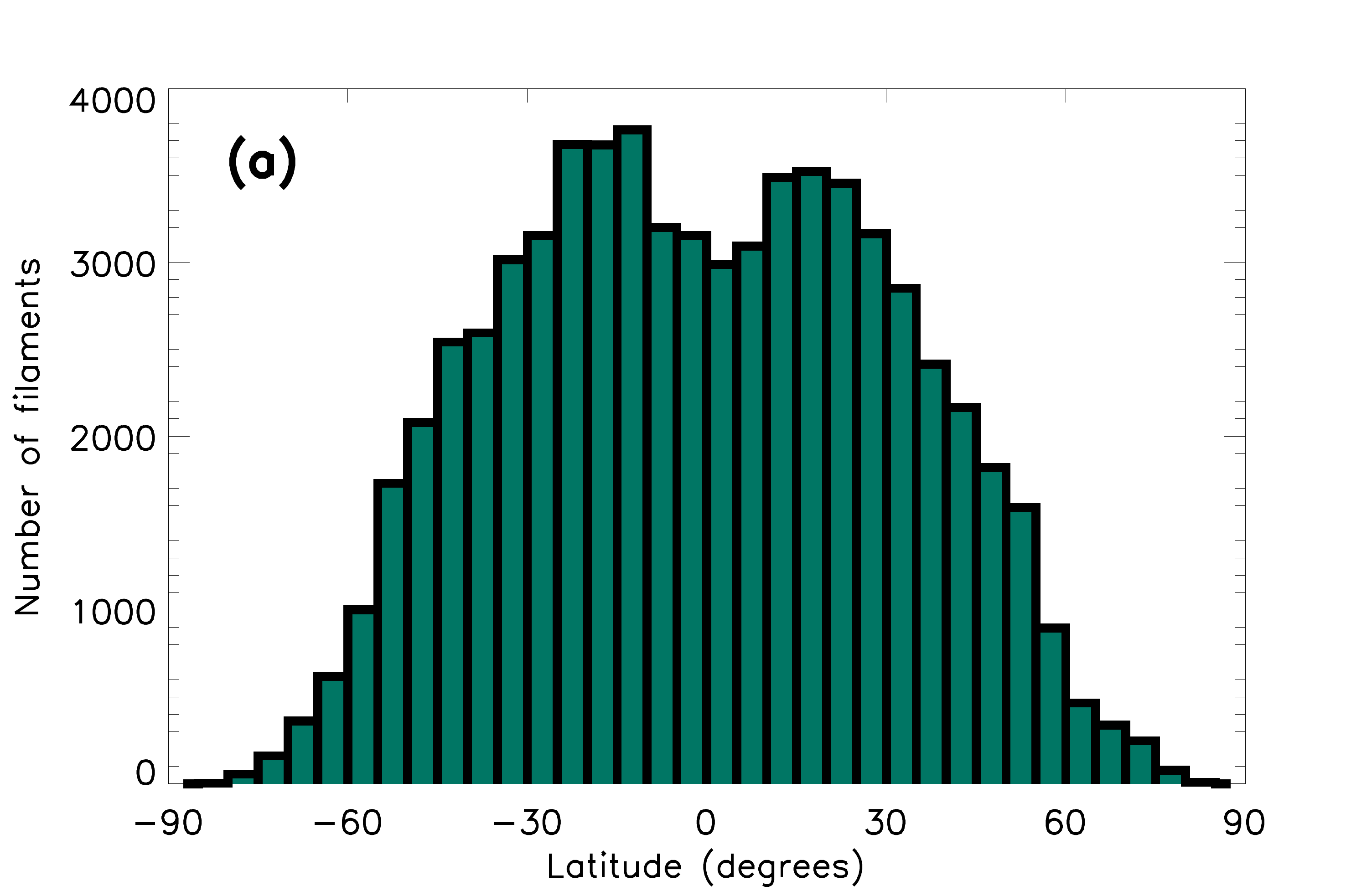}
  \end{minipage}
  \hfill
  \begin{minipage}[b]{0.4\textwidth}
    \includegraphics[width=\textwidth]{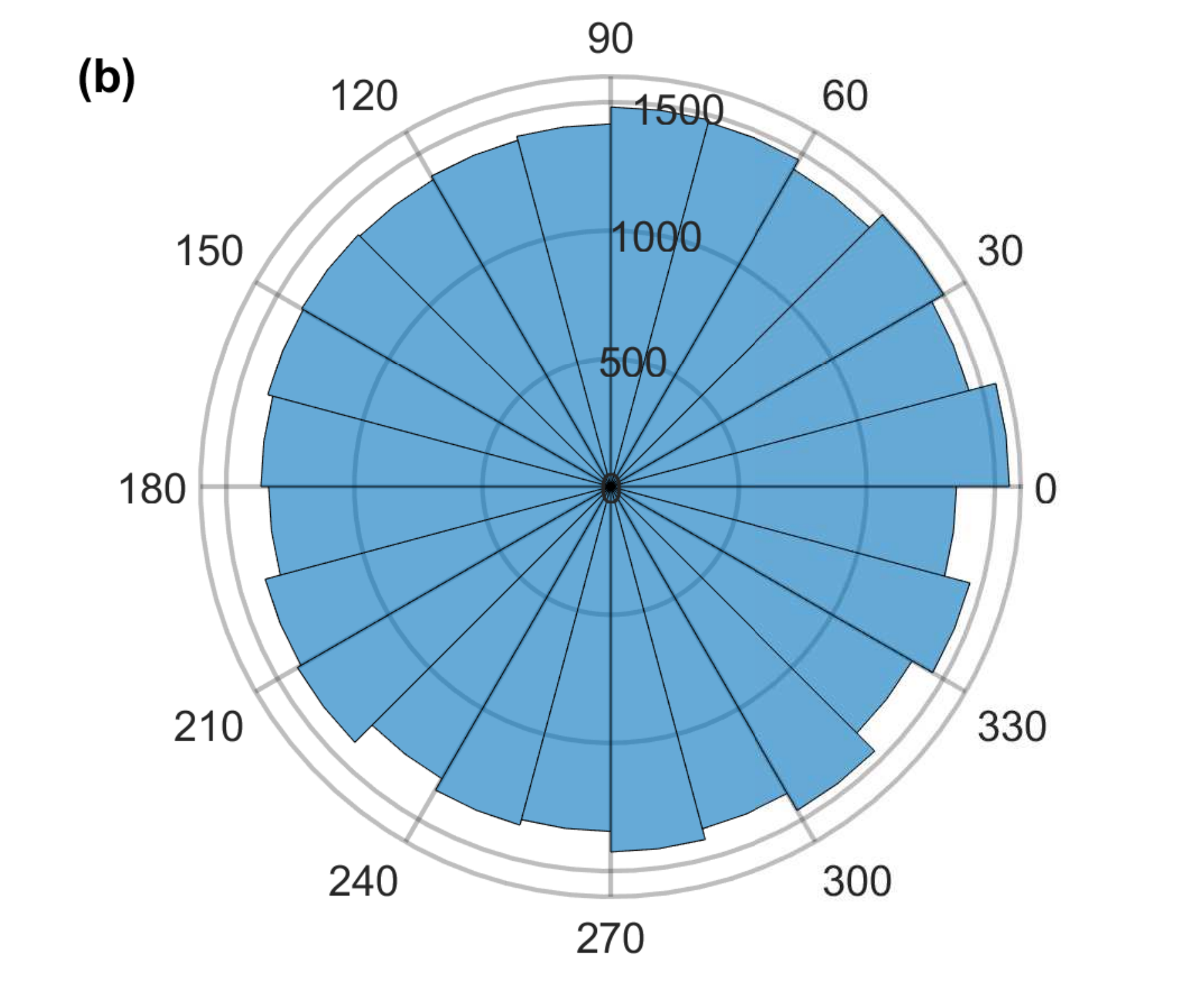}
  \end{minipage}
  \caption{In panel (a) the histogram represents the latitudinal distribution of filaments' geometric centers. The distribution shows a bimodal nature. In panel (b) the histogram depicted in a pie chart shows no longitudinal preference of filaments' geometric centers. }
\label{hist}  
\end{figure}

The histogram in Figure \ref{hist}(a) reveals that the latitudinal distribution of the geometric centers of filaments exhibits bimodal nature, with two peaks between $10^{\circ}$ to $30^{\circ}$ degree in both the hemispheres. The high concentration at low latitudes indicates that filaments are closely associated with the magnetic activity occurring around the active region belts in both the hemispheres. These results are consistent with the earlier findings by \cite{2015ApJS..221...33H} and \cite{2016SoPh..291.1115T} who have used different data sources. We further show that the longitudinal positions of filaments' centroids are evenly distributed over 360$^{\circ}$ through Figure \ref{hist}(b).

\subsubsection{Variation of filaments' length}
\begin{figure}[htb!]
\centering
\includegraphics[scale=0.9,angle=90]{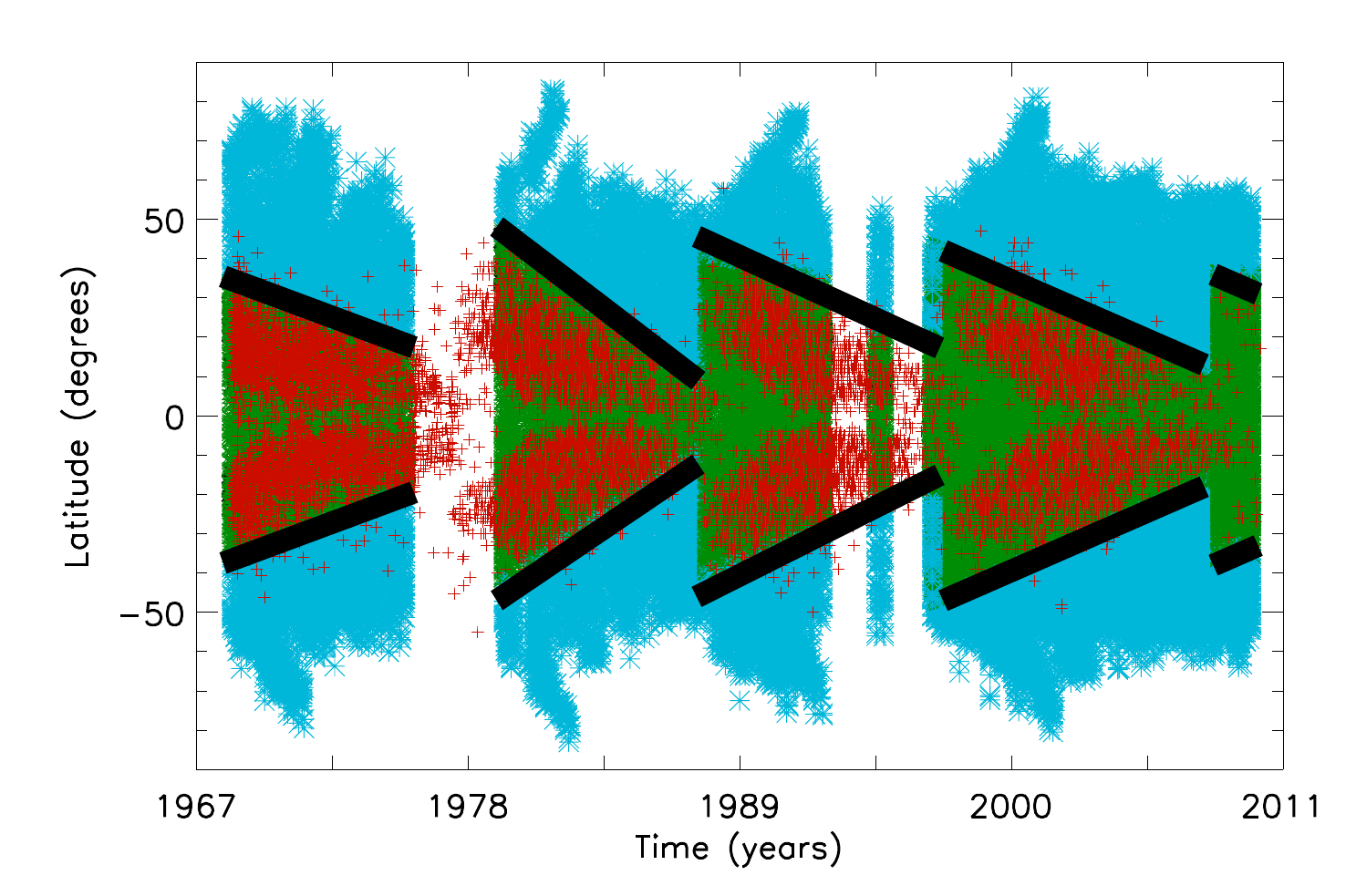}
\caption{Classification of filaments based on their relative latitudinal positions with respect to sunspot activity belt (marked by red `$+$' symbol). There are two classes: activity belt associated (\textit{AA}-) filaments and activity belt unassociated (\textit{UN}-) filaments. According to our convention, filaments with their geometric centers falling within the black lines are considered to be \textit{AA}-filament (marked by green asterisks) and rest are \textit{UN}-filaments (represented by blue asterisks).}
\label{fig4}
\end{figure}

We consider total filament length a better indicator of the solar magnetic activity as compared to the total number of filaments present in any individual Carrington map -- because the same number of filaments but with greater total length indicates the presence of higher magnetic flux content on the solar surface. We measure a quantity, $L_{total}$, the total filament length corresponding to a certain Carrington map by summing the length of all filaments present in that particular map, which shows a cyclic variation that is in phase with the 11-year solar cycle. To examine different origins of filament formation, we further sub-categorize the filaments into two classes depending on the association of their geometric center with the sunspot activity-belts where the positions of sunspots are determined from RGO-USAF/NOAA database. We define filaments with their geometric center falling within the pair of black lines (see, Figure \ref{fig4} in both the hemispheres as activity-belt associated filaments (\textit{AA-}, hereafter) and their cumulative length as $L_{AA}$. Filament centers situated outside of these black lines are considered to be activity-belt unassociated (\textit{UN-}) filaments, and their cumulative length is $L_{UN}$. The \textit{AA-}filaments and \textit{UN-}filaments are represented in Figure \ref{fig4} by green and blue asterisks, respectively. The emergence of sunspots along the activity belt can naturally influence the formation of \textit{AA}-filaments on a smaller scale; whereas the \textit{UN}-filaments can be considered to be predominantly governed by the large-scale structures of the magnetic field. 

\begin{figure}[htb!]
\centering
\includegraphics[scale=0.9,angle=90]{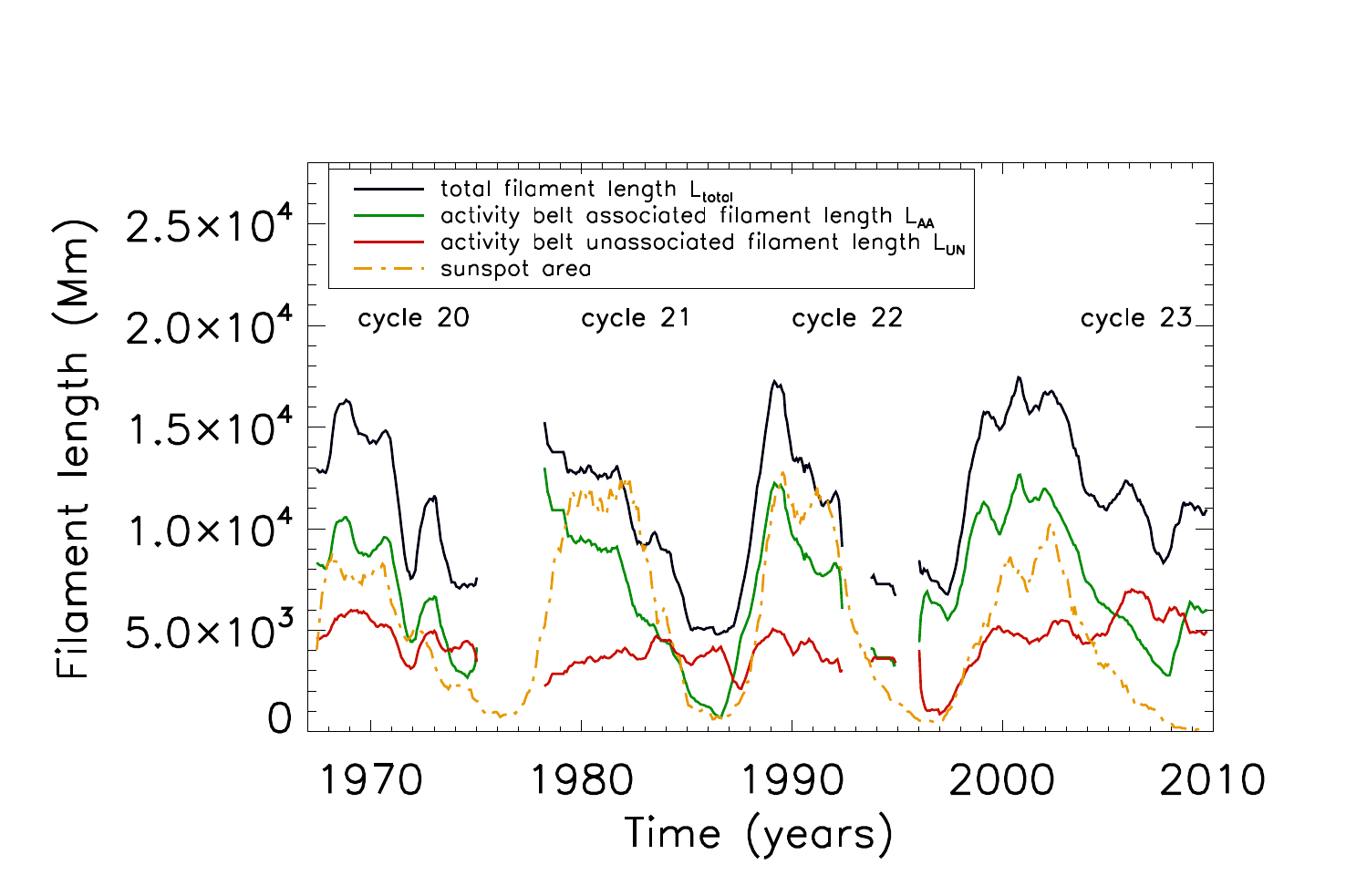}
\caption{Variation of filaments' length with time. The black curve shows modulation of the total filament length ($L_{total}$), exhibiting a notable cyclic nature which is almost in phase with the active region area variation depicted by the orange line. The green and red curves depict time evolution of \textit{AA}-filaments ($L_{AA}$) and \textit{UN}-filaments ($L_{UN}$), respectively during solar cycles 20--23. We note that all these variables shown in this figure are obtained after implementing thirteen-months running mean on the monthly data.}
\label{fig5}
\end{figure}

Figure \ref{fig5} depicts the variation of $L_{total}$ (black curve), $L_{AA}$ (green curve) and $L_{UN}$ (red curve) during cycles 20 to 23 in comparison to the active region (AR) area variation (denoted by the orange curve). It is evident that the \textit{AA}-filaments are the primary contributors to the total filament length and they display a prominent solar-cycle-like modulation. The complex magnetic field distribution throughout the activity belt creates a favorable environment for \textit{AA}-filament formation. However, during cycles 20, 21, 22 and 23 minima (around the year 1974, 1986, 1995 and 2008, respectively), $L_{UN}$ exceeds or becomes comparable with $L_{AA}$; which demonstrates that during solar minimum filaments mostly originate from the large-scale structures. In general, the filament length increases quite steeply during the rising phase of solar cycles compared to their descending phase. We speculate that the distinct slow decay rate of filament length throughout the descending phase of cycle 23 is coupled with the elongated minimum of that cycle. We cannot comment on this aspect regarding the other cycle minima due to the limited availability (or discontinuity) of data. In case of \textit{UN-}filaments, the cyclic modulation is less noticeable.

To quantify the inter-dependency between filament length and solar activity, we perform correlation analysis. The Spearman rank correlation coefficient between $L_{total}$ and AR area is 0.63 with a p-value of 0; whereas the degree of correlation increases in the case of \textit{AA-}filaments -- the rank correlation coefficient is 0.75 with p-value 0. However, the high latitude \textit{UN-}filaments do not correlate with AR area. \\
\subsubsection{Statistics of filaments' tilt angles}
\begin{figure}[htb!]
\centering
\includegraphics[scale=0.48,angle=0]{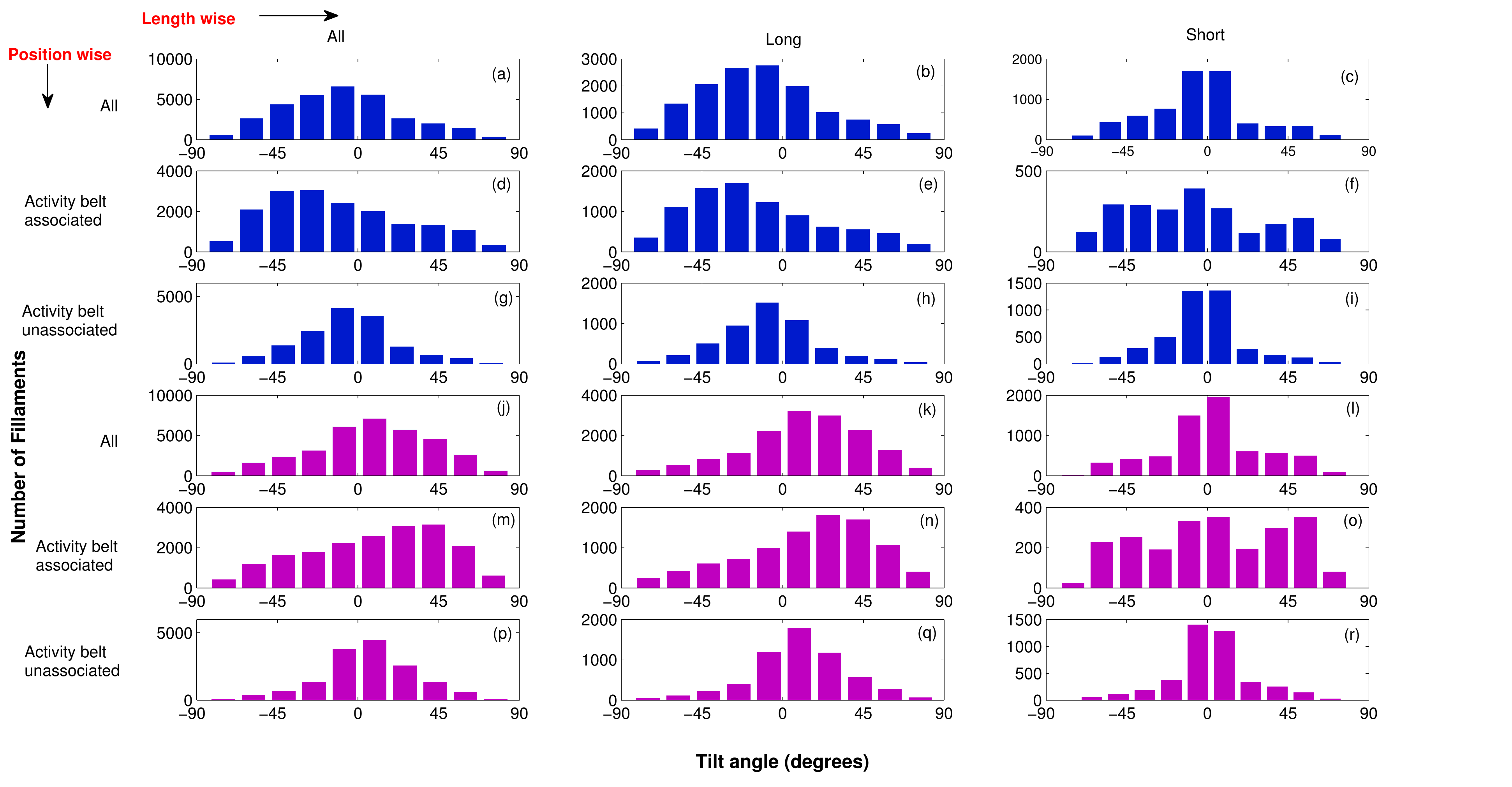}
\caption{Frequency distribution of tilt angles of filaments is plotted for different length and latitude criteria in the two hemispheres. Sub-figures (a) to (i) depict distributions of filament tilt angles in the northern hemisphere, whereas sub-figures in the last three panels [(j) to (r)] represent the same in the southern hemisphere. Sub-figure (a) corresponds to the distribution of all filaments in the northern hemisphere. Sub-figures (b) and (c) show the tilt angle distribution in the northern hemisphere for long ($>$ 52 Mm) and short ($<$ 28 Mm) filaments, respectively. Sub-figure (d) represents the distribution of \textit{AA-}filaments in the northern hemisphere, which is further sub-categorized in long and short filaments, as shown in (e) and (f). Tilt angle distribution associated with \textit{UN-}filaments are represented in sub-figures (g)-(i) with similar long-short length segregation. Sub-figures (j)-(r) outline the tilt angle frequency distribution under the same length and latitudinal classification for filaments present in the southern hemisphere. From these series of figures it is apparent that majority of the northern hemispheric filaments have negative tilt angles and southern hemispheric filaments exhibit dominance of positive tilt.}
\label{fig6}
\end{figure}

We determine the tilt of a particular filament by identifying the angle its axis forms with the local latitude. Each filament is fitted with a suitable straight line to deduce its axis and associated tilt angle. We presume that a thorough investigation of tilt angle distribution can shed light on the origin of filament formation. Figure \ref{fig6} consists of a series of histograms representing frequency distribution of filaments' tilt angles in the two hemispheres with different latitude and length criteria. In our study, filaments with their length varying from 0 to 28 Mm (i.e., mean$-{\sigma}$/2) are regarded as short, and filaments with the length greater than 52 Mm (i.e., mean$+{\sigma}$/2) are considered to be long.
In general, the filaments form along the PILs on the solar surface. Since the PIL associated with a bipolar Active Region (AR) separates positive and negative polarities, it is expected to be oriented perpendicular to the line joining the centers of the two spots of that particular AR. According to Hale's polarity law \citep{1919ApJ....49..153H}, the sunspot should have negative and positive tilt angles in the northern and the southern hemispheres, respectively. Therefore tilt angles of the filaments linked with ARs (those following Joy's law) should be positive in the northern hemisphere and negative in the southern hemisphere. The first three panels of Figure \ref{fig6} [images (a) to (i)] represent the frequency distributions of tilt angles in the northern hemisphere. Figure \ref{fig6}.(b) shows that longer filaments are prevalent. While plotting the histograms of tilt angle distribution of \textit{AA}-filaments under different classes, (i.e., long and short), we find that they are predominantly long with negative tilts [see, Figure \ref{fig6}.(d) - (f)]; hence they are not closely associated with the ARs. This result leads us to consider that long filaments form along the PILs which originate from the large-scale magnetic structure on the solar surface. Surprisingly, the short \textit{AA}-filaments [see, Figure \ref{fig6}.(f)] also have primarily negative tilt angles, which is contrary to our expectation of typical intra-AR filament orientation. There might be two possible explanations: first, a filament can form between two different ARs in which case tilt angle alignment is precisely opposite to an intra-AR filament, as described in \cite{2001ApJ...558..872M}; and second, these small \textit{AA}-filaments are generated from the same large-scale structures hosting long filaments. Since McIntosh database lacks detailed magnetic field information, we cannot conclusively determine the origin of these short \textit{AA}-filaments. Evidently, \textit{UN}-filaments mostly have negative tilt angles [Figure \ref{fig6}.(g) - (i)] as they are bound to originate from the large-scale structures. The last three panels [images: (j) to (r)] of Figure \ref{fig6} consist of histograms depicting the tilt angle distribution in the southern hemisphere. Irrespective of the length or positional classification, the tilt angles of the southern hemispheric filaments are predominantly positive. These observations confirm that large-scale magnetic structures primarily govern filament formation. The rest of the filaments with their length varying in the range of mean $\pm{\sigma}$/2 also show dominant negative (and positive) tilt angle in northern (and southern) hemisphere regardless of their latitudinal positions concerning the sunspot activity belt. Utilizing the positional information of sunspots present in the Carrington maps, We have simultaneously identified filaments with their geometric centers falling within ten degrees radii of any sunspot, and characterized them as sunspot-associated filaments. In the northern and southern hemispheres, the percentage of short sunspot-associated filaments with negative and positive tilt are 62$\%$ and 58$\%$, respectively, consistent with the results obtained in case of short \textit{AA}-filaments. Since the McIntosh database only provides positional information, investigating the chirality of filaments \citep{1998SoPh..182..107M,2007SoPh..245...87Y,2008SoPh..247..103Y,2009SoPh..254...77Y} was beyond our scope. The above analyses corroborate that distribution of tilt angles over latitude is not random \citep{2016SoPh..291.1115T} but follows systematic rules. 

\begin{figure}[htb!]
\centering
\includegraphics[scale=0.9,angle=90]{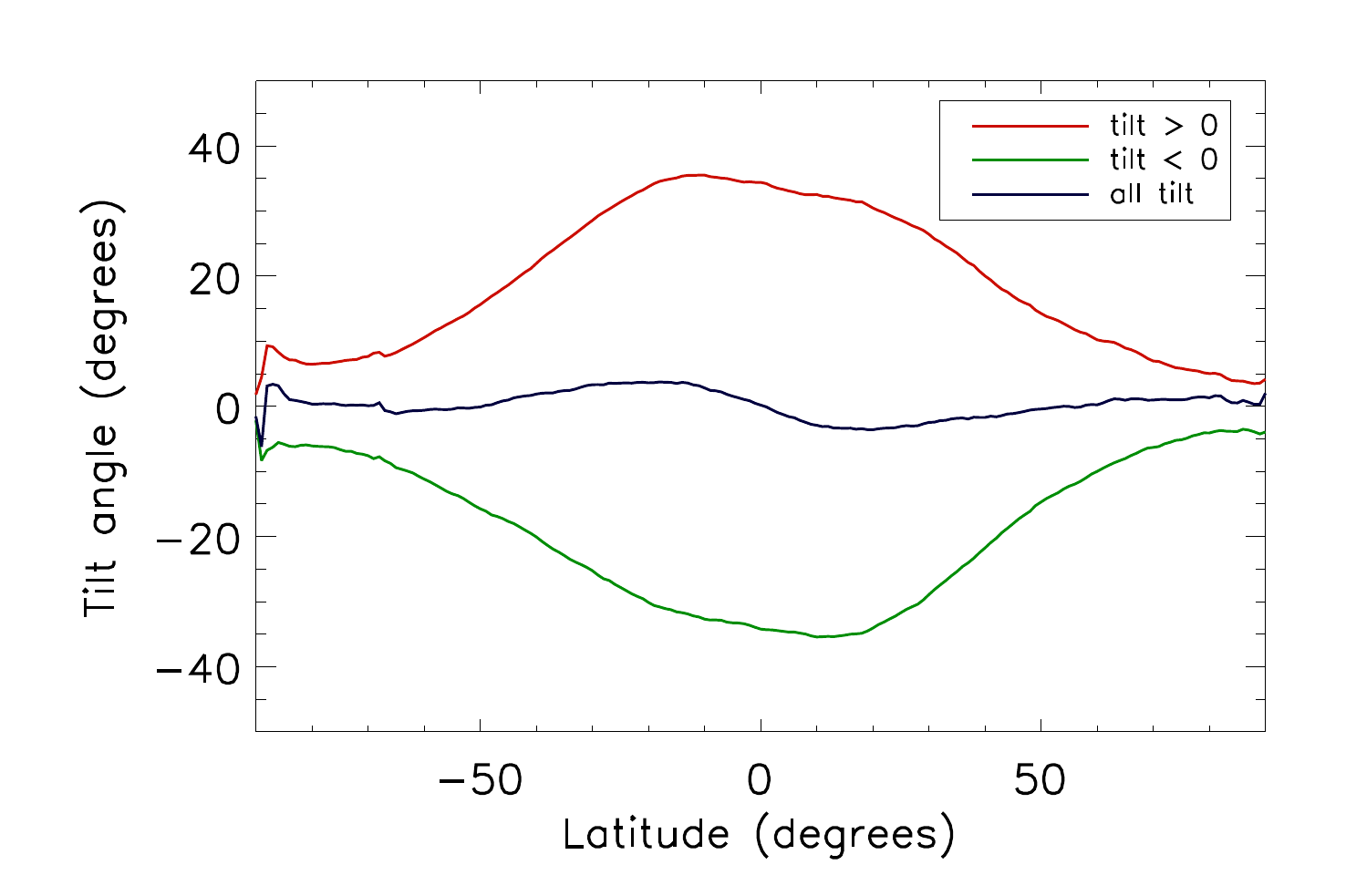}
\caption{Average tilt angle variation of filaments with latitude. Red and green curves depict the variation of average positive and negative tilt angle of filaments with latitude, respectively. The blue curve shows the modulation of the averaged cumulative tilt angle (signed) with latitude.}
\label{tiltlat}
\end{figure}

We further perform a Rayleigh test of uniformity on the tilt angle data of filaments belonging to the northern and the southern hemispheres separately. The Rayleigh test evaluates whether the tilt angles are equally scattered over the full range of  --90$^{\circ}$ to 90$^{\circ}$ based on circular statistical data analysis. In case of the northern hemispheric filaments' tilt angles, the Rayleigh Z statistic value is 0.5 with a p-value of zero, whereas, the test statistics value is 0.49 with a p-value of zero for those present in the southern hemisphere. The test results reject the null hypothesis of uniformity thus indicating that the tilt angles of the filaments are not evenly distributed over \ --90$^{\circ}$ to 90$^{\circ}$. In Figure \ref{tiltlat}, we study the variation of average tilt angle (denoted by a blue curve), which is primarily negative in the northern and positive in the southern hemisphere -- validating our earlier findings in the analyses of tilt angle distribution. In the same figure, the modulation of average positive (red curve) and negative tilt angle (blue curve) with latitudes are depicted. We notice that the average signed tilt angle is maximum near the equator and it decreases as we move to higher latitudes. It happens because, with increasing latitude, the PILs hosting the filaments become more parallel to the equator.\\ \\ \\
\subsection{Properties of polarity inversion line}
The PILs are the lines separating two opposite magnetic polarities on the solar surface. We measure the length of every PIL by using the same relation described in equation (1), and finally calculate the total length of PILs for each of the 442 Carrington maps (April 1967 to July 2009) by adding their individual lengths.

\subsubsection{Variation of polarity inversion line length}
\begin{figure}[htb!]
\centering
\includegraphics[scale=1,angle=90]{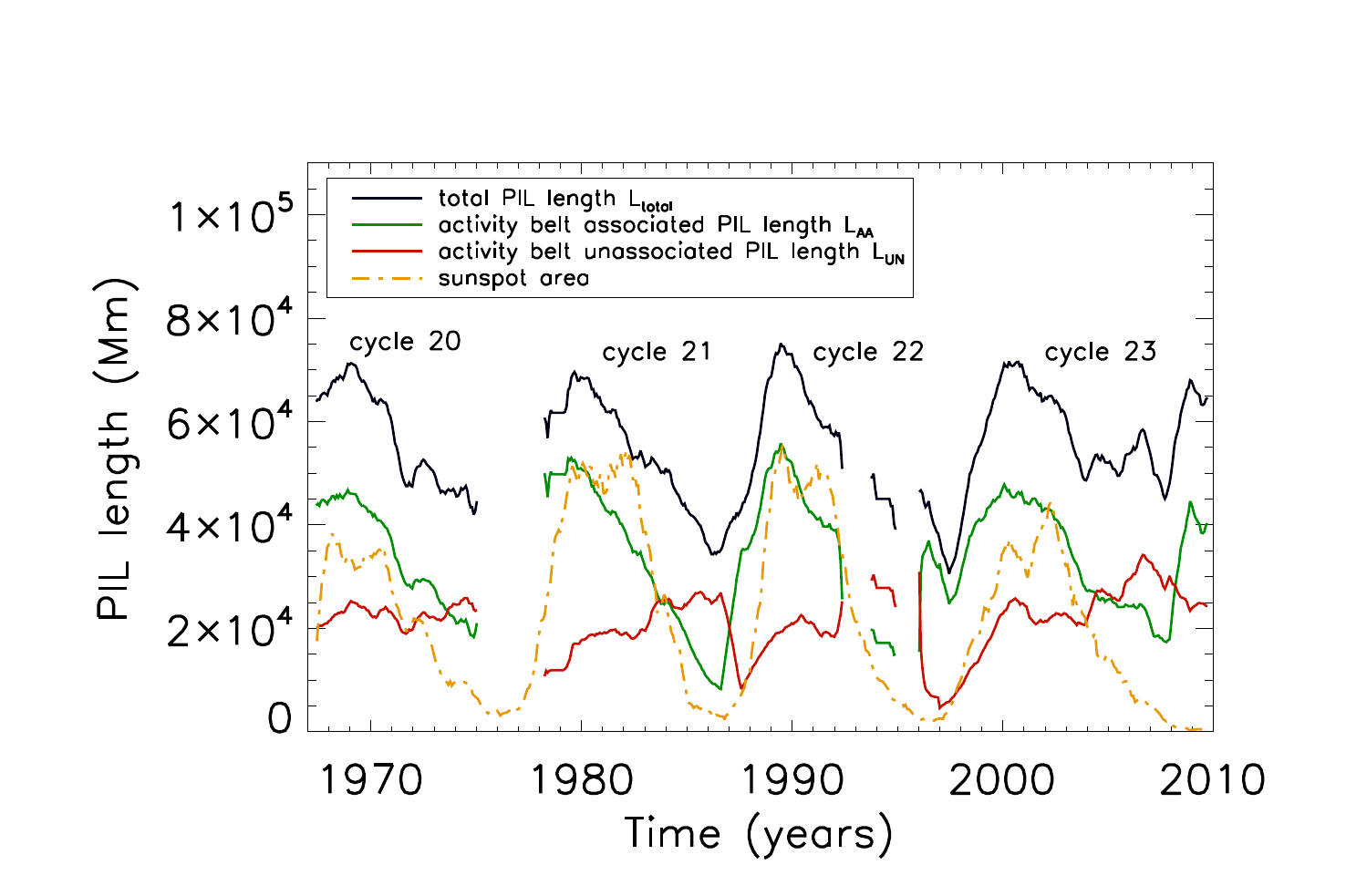}
\caption{Total PILs length (denoted by the black curve) varies during a solar cycle and shows a positive correlation with the sunspot area, the evolution of which is represented by the orange curve. The green and red curves show length variation of \textit{AA-}PILs and \textit{UN-}PILs, respectively during the last four solar cycles. Consistently, we applied thirteen-months running mean on every variable presented in this figure.}
\label{pil_ln}
\end{figure}

The time variation of the total length of the PILs (represented by the solid black line in Figure \ref{pil_ln}) shows a strong correlation with the cycle of sunspot area (the orange line). The Pearson linear correlation coefficient is 0.61 with p-value $ 6.37 \times 10^{-46} \hspace{0.1cm}$. The total length is primarily influenced by the activity-belt associated (\textit{AA-}) PILs (solid green line), which we measure by adding the PILs length belonging to PILs within the region enclosed by the pair of black lines in Figure \ref{fig4}. We believe that the complex distribution of PILs around ARs amplifies the overall length in lower latitudes. As the number of ARs drops significantly during cycle minima, we observe a decrease in \textit{AA-}PIL length during those periods. We notice a general growth in the length of activity-belt unassociated PILs during solar minima (through the years, 1974--1975, 1984--1987, 1994--1995 and 2005-2008) -- reasonably similar to the variation of filament length. However, we encounter an unexpected rise in the PIL length since 2008, which cannot be explained by our current understanding of PIL formation. Since filament always forms over the PILs, a strong correlation exists between PIL and filament length with a Pearson linear correlation coefficient 0.74 with p-value $ 2.25 \times 10  ^{-77}$.

\subsection{Properties of coronal hole}
Apart from filaments and PILs, the Carrington maps in the McIntosh archive also stores information of the coronal holes appearing on the solar disk. Our primary objective is to investigate the time evolution of coronal hole area over the solar cycle. Total 442 Carrington maps from the McIntosh archive are used to acquire coronal hole area information. Apart from three aforementioned significant data gaps in this database (see, Section 2), coronal holes are absent from all Carrington maps during the period of April 1967 -- November 1973.
We compute the coronal hole area using the following relation 
\begin{equation}
 A = \sum_{n}R_{\odot}^2 \hspace{0.07cm}Cos \theta \hspace{0.07cm} \delta \theta \hspace{0.07cm} \delta \phi
\end{equation}
\noindent where A is coronal hole area, $R_{\odot}$ is radius of the Sun, the symbols $\theta$ and $\phi$ represent the latitude and the longitude of a particular pixels and $n$ is the total number of pixels associated with the coronal hole. The quantities $\delta \theta$ and $\delta\phi$ are the latitudinal and longitudinal coverage of the pixel in consideration, associated with the structure of the coronal hole.

\subsubsection{Coronal hole area variation}
\begin{figure}[htb!]
\centering
\includegraphics[scale=1,angle=90]{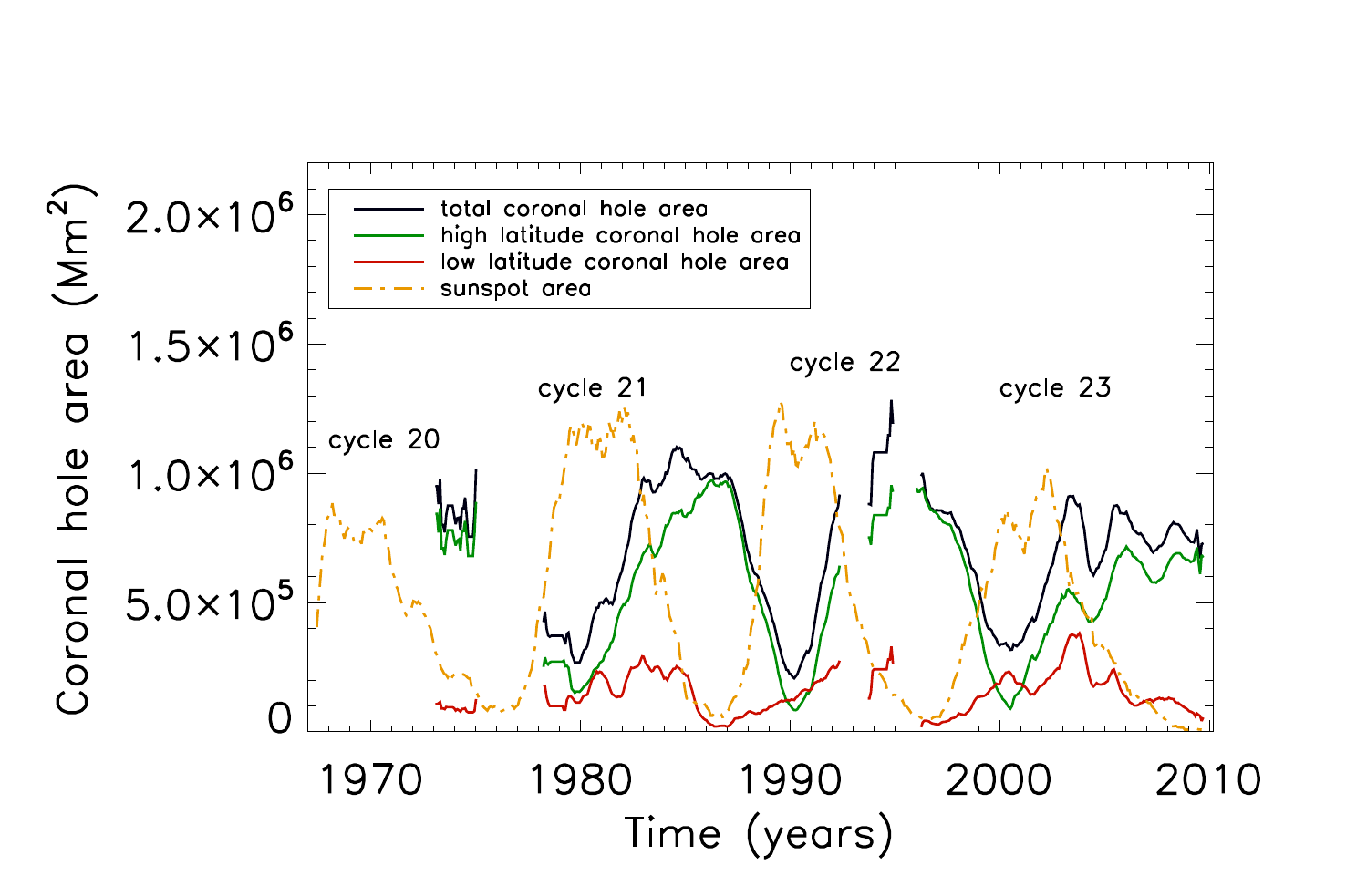}
\caption{The black curve represents the variation of coronal hole area during solar cycles 20--23. The area corresponding to high latitude ($>$ 40 $^{\circ}$) coronal holes depicted by the green curve is the primary contributor to the total coronal hole area which shows an overall negative correlation with the sunspot area variation denoted by the orange curve. The total area variation associated with the low latitude coronal holes is depicted by the red curve. All these quantities are obtained after applying thirteen-months running average on the monthly data.}
\label{ch_ar}
\end{figure}

In Figure \ref{ch_ar}, the time variation of total coronal hole area associated with individual Carrington maps along with the sunspot area is depicted. We define coronal holes present above 40$^{\circ}$ latitudes as high latitude coronal holes. The time evolution of the area associated with these high latitude coronal holes is also shown in the same figure by a green curve. It is quite apparent that coronal holes primarily appear at higher latitudes throughout the major period of solar cycles. The high latitude coronal hole area reaches its peak value during solar minima (e.g., cycle 21 minimum: the year 1985, cycle 22 minimum: the year 1995, although for cycle 23 in the year 2004), and it decreases as the sunspot cycle approaches its maximum activity. In contrary, the low latitude coronal holes (depicted by the red curve) appear predominantly during the solar cycle and generally disappear during sunspot minimum. The correlation analysis between high latitude coronal hole area and sunspot area shows a negative correlation; the Pearson linear coefficient is $-$0.58 with p-value $6.8 \times 10^{-34}$. However, the low latitude coronal holes do not possess any correlation with sunspot area, resulting in overall decrease in the correlation value between the total coronal hole area and sunspot area, such that Pearson correlation coefficient is $-$0.43 with p-value $1.3 \times 10^{-17}$. Coronal holes always appear over the open magnetic field lines which are extended from the photosphere to solar corona and mostly populated beyond 40$^{\circ}$ latitudes in both the hemispheres. The high latitude open field lines present near the polar regions are governed by the global dipolar magnetic configuration of the Sun. The amplitude of this dipole becomes minimum during a sunspot cycle maximum. Therefore the prominent negative correlation between high latitude coronal hole area and sunspot area is quite apparent in this context.\\ \\

\subsection{Hemispheric asymmetry in filament length, PIL length, and coronal hole area}

\begin{figure}[htb!]
\centering
\includegraphics[scale=0.9,angle=90]{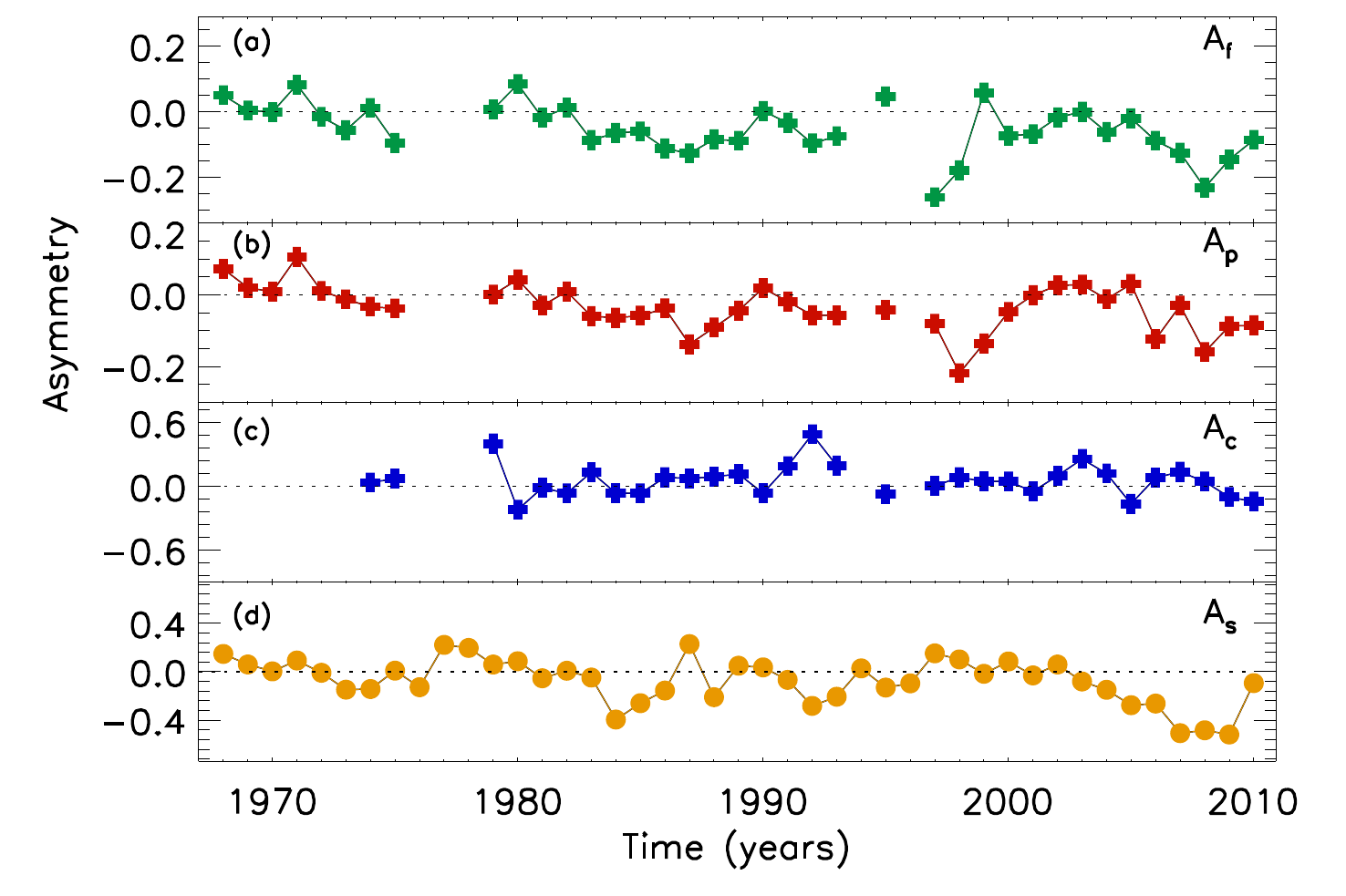}
\caption{The series of images present the time variation of yearly averaged north-south asymmetry index of various observed solar features. Hemispheric asymmetry in (a) filaments length, (b) PILs length, (c) coronal hole area and (d) sunspot area are depicted by green, red, blue and orange curves, respectively.}
\label{asymm}
\end{figure}

The solar magnetic activity is observed to be somewhat asymmetric in the two hemispheres, which inherently introduces north-south asymmetry in different magnetic features appearing on the solar photosphere and corona. We define the asymmetry in the sunspot area which reasonably represents the magnetic flux input on the solar surface by a quantity $A_s$, 

\begin{equation}
A_{s}=\frac{N_{s}-S_{s}}{N_{s}+S_{s}},
\end{equation}

\noindent where $N_{s}$ and $S_{s}$ are the total area of the sunspots those emerged during each year of the observational period in the northern and the southern hemispheres, respectively. The time variation of $A_s$ during solar cycles 20 to 23 is depicted by the orange curve in panel (d) of Figure \ref{asymm} showing a surplus of magnetic activity in the southern hemisphere. In the following sections, we explore the hemispheric asymmetry of different observables in the McIntosh database and their degree of correlation with the asymmetry present in sunspot area. 

\subsubsection{North-south asymmetry of filaments}
Earlier studies \citep{2003NewA....8..655L,2010NewA...15..346L,2015RAA....15...77K,2015ApJS..221...33H} have reported that the filament number possesses hemispheric asymmetry, and it changes with time. We define asymmetry in filament length by a quantity $A_f$,

\begin{equation}
A_{f}=\frac{N_{f}-S_{f}}{N_{f}+S_{f}},
\end{equation}

\noindent where $N_{f}$ and $S_{f}$ are the total filament length in northern and southern hemispheres, respectively. Naturally, if $A_{f} > 0$, the total filament length in the northern hemisphere dominates that in the southern hemisphere. We find $A_{f}$ to fluctuate over the time of observation with a prominent inclination towards the southern hemisphere (see, panel (a) of Figure \ref{asymm}) during the period of observation. Only during a few short periods (for example, the years 1968, 1971, 1980, 1995, 1999), we observe filaments belonging to the northern hemisphere dominate those in the south. The hemispheric asymmetries in total filament length shows a weak correlation with the sunspot area with a linear correlation coefficient of 0.32 with p-value 0.0493.

\subsubsection{North-south asymmetry of polarity inversion line}
The development of PILs is entirely regulated by the magnetic activity on the solar surface, which usually exhibits hemispheric asymmetry. Therefore, PILs are bound to show hemispheric preferences. We define normalized north-south asymmetry index for PIL ($A_{p}$) by the following formula 
\begin{equation}
A_{p}=\frac{N_{p}-S_{p}}{N_{p}+S_{p}},
\end{equation}
\noindent where $N_{p}$ and $S_{p}$ are the total length of PILs in northern and southern hemispheres, respectively. If $A_{p} > 0 $, the northern hemisphere shows dominance over the southern hemisphere regarding the total length of PILs. Figure \ref{asymm}(b) depicts the variation of yearly averaged north-south asymmetry index of the PILs. Similar to the hemispheric asymmetry existing in filament length, we observe a southern hemispheric dominance in the PILs length over the whole period of the available data. The hemispheric asymmetries in total PIL length shows a weak correlation with the sunspot area with a Spearman rank correlation coefficient of 0.30 with p-value 0.0719.
A correlation analysis between the hemispheric asymmetry found in filaments and PILs confirms that the origin of filaments is physically interconnected with the PILs. We find the Spearman rank correlation coefficient to be as high as 0.69 with a p value 2.92$\times 10^{-6}$.

\subsubsection{North-south asymmetry in coronal hole area}
We define the normalized asymmetry index of coronal hole area by 
\begin{equation}
A_{c}=\frac{N^c_{north}-N^c_{south}}{N^c_{north}+N^c_{south}},
\end{equation}
\noindent where $N^c_{north}$ and $N^c_{south}$ are the total area of coronal holes in northern and southern hemispheres, respectively. The index, $A_{c} > 0 $ implies that the total area of coronal holes in the northern hemisphere exceeds those in the southern hemisphere. The modulation of $A_{c}$ is shown in Figure \ref{asymm}(c), which mostly varies between $+/-$ 0.1 except during few years. Unlike the asymmetry in filaments or PILs length, $A_{c}$ fluctuates between positive and negative values throughout the time-span of observation, indicating no consistent hemispheric dominance. Furthermore, asymmetry in coronal holes is almost uncorrelated with the sunspot area asymmetry. 

\section{Conclusions}
We have explored different properties of filaments, PILs, and coronal holes during past four solar cycles, starting from the maximum of solar cycle 20 (April 1967) to the beginning of cycle 24 (July 2009) using the McIntosh database. Despite the presence of data gaps, our analysis of 442 Carrington maps spread over 36 years establishes some important trends. We find a cyclic behavior of total filament length with the time that is consistent with earlier reports \mbox{\citep{2015ApJS..221...33H,2016SoPh..291.1115T}}. The variation of the total PIL lengths with time shows a cyclic behavior similar to the solar cycle. Analyzing the PILs length in the backdrop of their association with the sunspot activity belt reveals that activity-belt associated (\textit{AA-}) PILs contribute more towards the cumulative trend. We find a negative correlation between high latitude coronal hole area and sunspot area. The high latitude ($>$ 40$^\circ$) coronal holes are the main contributors to their total area. This modulation of coronal hole area with time is understood in the context of the evolution of open magnetic field lines through different phases of a solar cycle. We further find that hemispheric asymmetry in the filament and PIL are weakly correlated with asymmetry in sunspot areas. However, the asymmetry in coronal hole area is found to be uncorrelated with sunspot area asymmetry suggesting processes other than (asymmetric) flux emergence may govern coronal hole asymmetry. 

Additionally, we classify filaments based on the latitudinal positions of their geometric centers to explore their connection with the photospheric magnetic environment. The activity-belt associated (\textit{AA-}) filaments contribute more towards the overall cyclic variation compared to the activity belt unassociated (\textit{UN-}) filaments -- thus implicating complex magnetic field distribution and flux emergence and cancellation within the sunspot activity belt. The latitudinal distribution of filaments is bimodal with peaks between $10^{\circ}$ to $30^{\circ}$ in both the hemispheres which is consistent with the overall latitudinal profile of magnetic flux emergence.

The butterfly-like spatio-temporal distribution of long filaments reveals a distinguishable feature -- the poleward migration of high-latitude filaments during solar maxima. Through quantitative analysis we find this to be co-temporal with the initiation of global, solar polar field reversal. We propose that high latitude filament migration could be a tracer of flux cancellation at the high-latitudes which leads to the reversal of the old cycle polarity.   

Average tilt angles of filaments decrease from the equator towards poles, consistent with earlier findings by \mbox{\cite{2016SoPh..291.1115T}, \cite{2017ApJ...849...44C} and \cite{2015ApJS..221...33H}}. The tilt angle statistics show an unexpected dominance of negative tilt angle in the northern hemisphere and positive tilt angle dominance in the southern hemisphere. The trend remains similar even when we segregate the filaments depending on their association with the sunspot activity belt. Based on this we conclude that filament tilt angle distribution is governed by the large-scale magnetic structures on the solar surface and not the orientation of small-scale intra or inter active region magnetic fields. In summary, our work has established the substantial dependency of filament formation on large-scale surface magnetic field distribution through detailed analyses. 

\section*{Acknowledgments}
We are grateful to Patrick S. McIntosh, NOAA Space Environment Laboratory (1964), for the creation of the hand-drawn Carrington maps using both satellite and ground-based observation. We also thank the team of McIntosh Archive (McA) project (a Boston College/NOAA/NCAR collaboration, funded by the NSF), NOAA National Centers for Environmental Information for preserving and making the open digitized McIntosh archive. Wilcox Solar Observatory polar field data used in this study was obtained via the web site \url{http://wso.stanford.edu} courtesy of J.T. Hoeksema, whereas the MWO polar field data were downloaded from the solar dynamo data-verse \url{https://dataverse.harvard.edu/dataverse/solardynamo}, maintained by Andr\'{e}s Mu\~noz-Jaramillo. We acknowledge utilization of data from the Royal Greenwich Observatory/USAF-NOAA active region database compiled by David H. Hathaway \url{(https://solarscience.msfc.nasa.gov/greenwch.shtml)}. R.M. acknowledges funding from University Grants Commission, Government of India. P.B. acknowledges funding by CEFIPRA/IFCPAR through grant 5004-1. D.N. acknowledges the Indo-US Joint Center Programme grant IUSSTF-JC-011-20 for facilitating interactions that inspired this research. We thank Sarah E. Gibson for helpful discussions related to this work. We are thankful to the anonymous reviewer for valuable suggestions which helped us to improve the quality of our work.

\bibliography{references}
\bibliographystyle{aasjournal}

\end{document}